# minoHealth.ai: A Clinical Evaluation Of Deep Learning Systems For the Diagnosis of Pleural Effusion and Cardiomegaly In Ghana, Vietnam and the United States of America

*Darlington, Akogo[1]; Benjamin Dabo Sarkodie[2], Issah Abubakari Samori[1]; Bashiru Babatunde Jimah[3]; Dorothea Akosua Anim[4]; Yaw Boateng Mensah [2];*

[1] minoHealth AI Labs, Accra, Ghana.

[2] University of Ghana Medical School, Department of Radiology, Accra, Ghana.

[3] University of Cape Coast, Department of Medical Imaging, Cape Coast, Ghana.

[4] Korle-Bu Teaching Hospital, Radiology Department, Accra, Ghana.

**ABSTRACT**

*A rapid and accurate diagnosis of medical conditions like cardiomegaly and pleural effusion is of the utmost importance to reduce mortality and medical costs. Artificial Intelligence has shown promise in diagnosing medical conditions. With this study, we seek to evaluate how well Artificial Intelligence (AI) systems, developed my minoHealth AI Labs, will perform at diagnosing cardiomegaly and pleural effusion, using chest x-rays from Ghana, Vietnam and the USA, and how well AI systems will perform when compared with radiologists working in Ghana. The evaluation dataset used in this study contained 100 images randomly selected from three datasets. The Deep Learning models were further tested on a larger Ghanaian dataset containing five hundred and sixty one (561) samples. Two AI systems were then evaluated on the evaluation dataset, whilst we also gave the same chest x-ray images within the evaluation dataset to 4 radiologists, with 5 - 20 years experience, to diagnose independently. For cardiomegaly, minoHealth.ai systems scored Area under the Receiver operating characteristic Curve (AUC-ROC) of 0.9 and 0.97 while the AUC-ROC of individual radiologists ranged from 0.77 to 0.87. For pleural effusion, the minoHealth.ai systems scored 0.97 and 0.91 whereas individual radiologists scored between 0.75 and 0.86. On both conditions, the best performing AI model outperforms the best performing radiologist by about 10%. We also evaluate the specificity, sensitivity, negative predictive value (NPV), and positive predictive value (PPV) between the minoHealth.ai systems and radiologists.*

## 1. INTRODUCTION

Across the globe, especially in LMICs, there is a severe shortage of radiologists leading to poor healthcare delivery, physician burnout and errors. A seamlessly integrated AI component within the imaging workflow would increase efficiency, reduce errors and achieve objectives with minimal manual input by providing trained radiologists with pre-screened images and identified features. Therefore, substantial efforts and policies are being put forward to facilitate technological advances related to AI in medical imaging. Almost all image-based radiology tasks are contingent upon the quantification and assessment of radiographic characteristics from images. These characteristics can be important for the clinical task at hand, that is, for the detection, characterization or monitoring of diseases.

### 1.1 Cardiomegaly

Cardiomegaly refers to an enlarged heart identified via an imaging test. It has become increasingly prevalent and carries a high mortality. Heart damage and other conditions can lead to an enlargement of heart. Conditions associated with cardiomegaly include congenital heart defect, cardiomyopathy (diseases of the heart muscle), damage from a heart attack, pericardial effusion, hypertension (high blood pressure), pulmonary hypertension, and anemia.

## 1.2 Pleural Effusion

Pleural effusion is the build-up of excess fluid between the layers of the pleura outside the lungs. The pleura are thin membranes that line the lungs and the inside of the chest cavity and act to lubricate and facilitate breathing. Normally, a small amount of fluid is present in the pleura. Pleural effusion can be caused by pneumonia, cancer, pulmonary embolism, kidney disease, inflammatory disease, heart failure, pulmonary embolism, cirrhosis, and post open heart surgery.

## 1.3 Challenges Facing Radiology

As pointed out in the study on AI for Radiology (Akogo, 2020), radiology is very important, however, there's a shortage of radiologists globally, especially in developing countries. Liberia, for example, only has about 2 radiologists (RAD-AID, 2017), whilst Ghana has less than 80 radiologists and Kenya has 200 radiologists (UCSF, 2015). And in the UK, only one-in-five trusts and health boards has sufficient number of interventional radiologists to run a safe 24/7 service to perform urgent procedures (Clinical Radiology UK Workforce Census Report, 2018) whilst their workload of reading and interpreting medical images has increased by 30% between 2012 and 2017. There's a need for scalable and accurate automated radiological systems. Deep Learning, especially Convolutional Neural Networks, is gaining wide attention for its ability to accurately analyze medical images, with the potential to help solve the shortage of radiologists.

## 1.4 Artificial Intelligence In Radiology

Also, as pointed out in the study on AI for Radiology (Akogo, 2020), the re-emergence of Artificial Intelligence (A.I) and Deep Learning, due to growth in computing power and data, has led to advancements in Deep Convolutional Neural Networks, which has allowed for breakthrough research and applications in Radiology. Artificial Intelligence and Deep Learning holds a lot of potential in Radiology. Artificial Intelligence can provide support to radiologists and alleviate radiologist fatigue. It can help in flagging patients who require urgent care to radiologists and physicians. Deep Learning could also help increase interrater reliability among radiologists throughout their years in clinical practice. A recent study found that the Fleiss' kappa measure of interrater reliability for detecting anterior cruciate ligament tear, meniscal tear, and abnormality were higher with model assistance than without it (Bien et al., 2018). Deep Learning has achieved performances comparable to humans and sometimes better. A recent study analyzed 14 research works done using Deep Learning to detect diseases via medical images, they found that on average, Deep Learning systems correctly detected a disease state 87% of the time – compared with 86% for healthcare professionals – and correctly gave the all-clear 93% of the time, compared with 91% for human experts (Liu et al., 2019). Deep Learning has performed as well as radiologists and sometimes better at detecting abnormalities like pneumonia, fibrosis, hernia, edema and pneumothorax in chest x-rays (Rajpurkar et. al, 2017). It has also been used to detect knee abnormalities via magnetic resonance (MR) imaging at near-human-level performance (Bien et. al, 2018). Researchers have also trained Deep Learning models that outperformed dermatologists at detecting skin cancer (Esteva et. al, 2017, Haenssle et. al, 2018).

## 2. MATERIALS & METHOD

### 2.1 Double Blind Study Dataset

The chest conditions we focused on are Cardiomegaly and Pleural Effusion. The evaluation set used in this study contained hundred (100) images randomly selected from three datasets. The images were either healthy or contained pleural effusion or cardiomegaly or both (Figure 1). Twenty (20) images were selected from the VinBig Data Chest X-ray dataset (Nguyen et al., 2022.), twenty one (21) images were selected from the Chexpert dataset (Irvin et al., 2019), and fifty nine (59) images were selected from the Euracare dataset, an in-house dataset collected from the Euracare Advanced Diagnostics and Heart Centre. The VinBig Data Chest X-ray dataset is a dataset of chest x-rays collected by the Vingroup Big Data Institute in Vietnam. The Chexpert dataset is also another large dataset of chest X-rays collected by the Stanford ML group in the United States of America. Lastly, the Euracare data is an in-house dataset collected by minoHealth.ai from Euracare Advanced Diagnostics and Heart Centre, a top-tier health institution in Accra, Ghana. These different data sources were considered because we wanted the dataset used for this study to represent a diverse demographic. 64% of samples in the evaluation dataset were scans of males and 36% were scans of females. The distribution of the dataset based on country is: 59% of the samples were Ghana, 20% from Vietnam, and 21% from USA. The experts that took part in this study are radiologists from the Euracare Advanced Diagnostics and Heart Centre and the Korle Bu Teaching Hospital. Two minoHealth.ai's models, that represent different machine learning techniques, were used in this study.

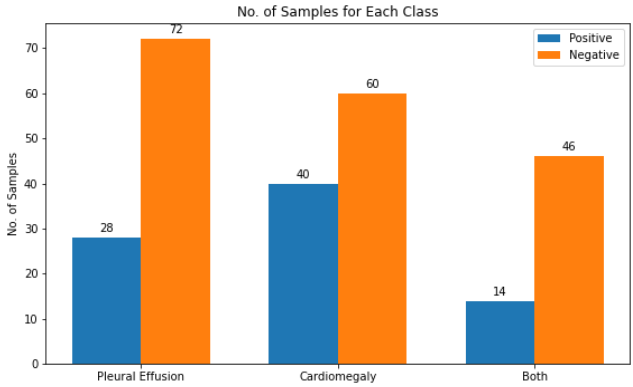

**Figure 1** - The bar chart shows the number of positive and negative samples for cardiomegaly and pleural effusion. It also shows the number of samples that have both disease conditions and the number of samples that have neither disease conditions.

## 2.2 Data collection and analysis

The images for the double-blind study were shuffled together and randomly named. The renamed images were distributed to each radiologist. The diagnosis of each radiologist was saved in an excel sheet for analysis. The Receiver Operating Characteristic (ROC) curve and Area Under ROC curve (AUC-ROC) for each radiologist and AI model were generated and compared across different demographics of the dataset. The AUC-ROC is a metric to tell how well an entity (AI system or radiologist) can differentiate between a diseased sample and a healthy sample (Hajian-Tilaki, 2013), and computing this metric for different demographics gives more information on how well a participating entity is able to differentiate between healthy and diseased samples across different demographics. The performance of the AI models and radiologists were also compared using the following metrics: Negative Predictive Value (NPV), Positive Predictive Value (PPV), Specificity, and Sensitivity (Table 1). Table 1 defines these metrics and provides their mathematical formulae in terms of True Positive (TP), True Negative (TN), False Positive (FP), and False Negative (FN). TP is the number of positive samples that were correctly diagnosed as positive and TN is the number of negative samples that were correctly diagnosed as negative. On the other hand, FP is the number of negative samples that were misdiagnosed as positive and FN is the number of positive samples that were misdiagnosed as negative.

**Table 1:** Table of additional metrics used to evaluate the performance of radiologists and AI models. The table also includes definition and mathematical formulae of the metrics (Steinberg et al., 2009)

| Metric | Definition | Mathematical formula |
|---|---|---|
| Positive Predictive Value (PPV) | The likelihood of a patient being positive (or diseased) when diagnosed as positive | $\frac{TP}{TP + FP}$ |
| Negative Predictive Value (NPV) | The likelihood of a patient being negative (or healthy) when diagnosed as negative | $\frac{TN}{TN + FN}$ |
| Specificity | The ability of a test model to identify negative or healthy samples | $\frac{TN}{TN + FP}$ |
| Sensitivity | The ability of a test model to identify negative or healthy samples | $\frac{TP}{TP + FN}$ |

## 3. RESULTS & DISCUSSION

This study involved six (6) participants - two proprietary minoHealth.ai's AI systems (named as minoHealth_model_1 and minoHealth_model_2) and four radiologists (named as Rad_1, Rad_2, Rad_3, and Rad_4). For comparison between radiologists and AI systems, the Receiver Operating Characteristic (ROC) Curve for each participating entity was generated and the area under the curves (AUCs) were computed. The AUC-ROC is a degree of separability and tells how well an AI system or a radiologist is capable of distinguishing a healthy scan from a diseased scan (Hajian-Tilaki, 2013). AUC-ROC is a value between 0 and 1, and the higher its value the better. On a high level, the ROC plots for each disease condition, cardiomegaly (Figure 2) and pleural effusion (Figure 3), were generated.

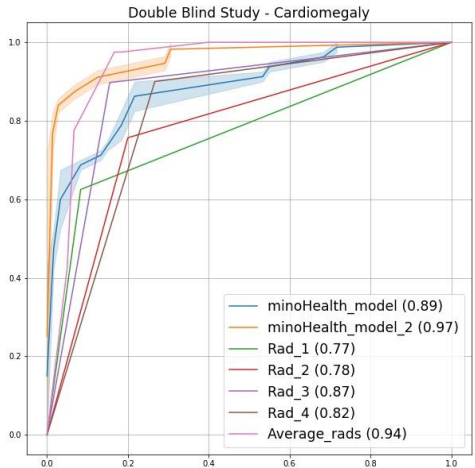

**Figure 2: ROC curves for AI models and radiologists for cardiomegaly. Average_rads represent the aggregate of predictions from all radiologists.**

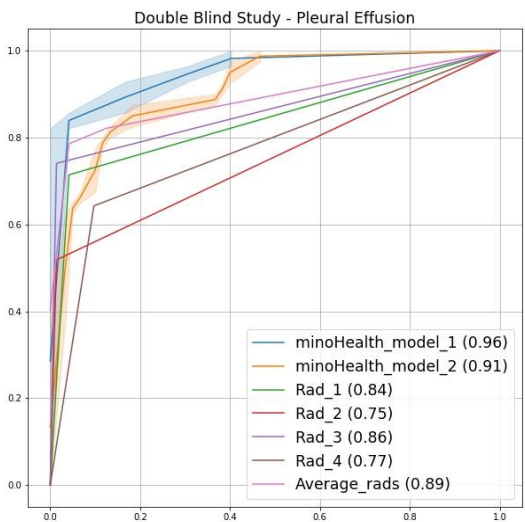

**Figure 3: ROC curves for AI models and radiologists for pleural effusion. Average_rads represent the aggregate of predictions from all radiologists.**

From Figures 4 and 5, both models outperform each radiologists on both disease conditions. For cardiomegaly, the models scored AUC-ROC of 0.89 and 0.97 while the AUC-ROC of individual radiologists ranged from 0.77 to 0.87 (Figure 2). For pleural effusion, the models scored 0.96 and 0.91 whereas individual radiologists scored between 0.75 and 0.86 (Figure 3). On both conditions, the best performing AI model outperforms the best performing radiologist by about 10%.

Also, it is worth noting that the AUC-ROC value varies from one radiologist to another. This implies that for some medical images, the different radiologists diagnose them differently and did not have an objective diagnosis. This reinforces the fact that there is some level of subjectivity and personal errors in diagnosing medical images by human experts (Tarkiainen et al., 2021). These individual subjectivities and personal errors is what has caused the different diagnosis by different radiologists. To compare the performances of our AI models against human experts without any level of subjectivity or personal errors, we aggregated the diagnosis of all radiologists by averaging the diagnosis of all radiologists. This is similar to multiple radiologists working together to diagnose an image. The aggregate of the radiologists' diagnoses (labeled Average_rads) scored AUC-ROC of 0.89 for pleural effusion (versus 0.96 and 0.91 by AI models) and 0.94 for cardiomegaly (versus 0.97 and 0.89 by AI models).

The performance of the AI models are comparable to, if not better than, the performance of the aggregate of radiologists. This implies that the performance of the AI models is comparable to multiple radiologists working together to diagnose an image. In regions like Sub Saharan Africa, where radiologists are scarce and are also overloaded with other clinical responsibilities (Mutala et al., 2020), solutions like the AI models will be of great utility. These solutions can achieve the performance of multiple radiologists working together to complement the efforts of radiologists and ease the burden on them.

To probe further into the performance of the AI models across different demographics, the ROC curves for specific demographics groups were generated. Gender-wise, the AI models were observed to perform better when diagnosing female chest x-rays than they were at diagnosing male chest x-rays (Figures 6-9).

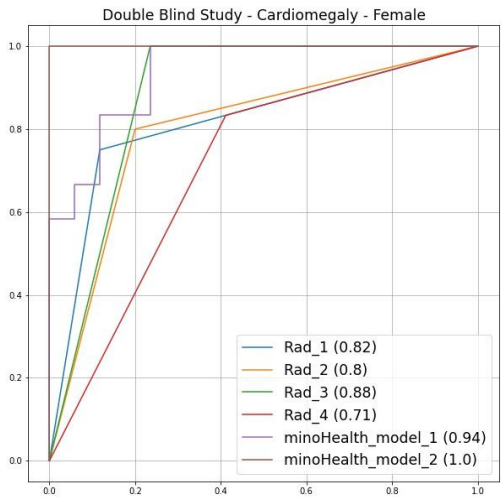

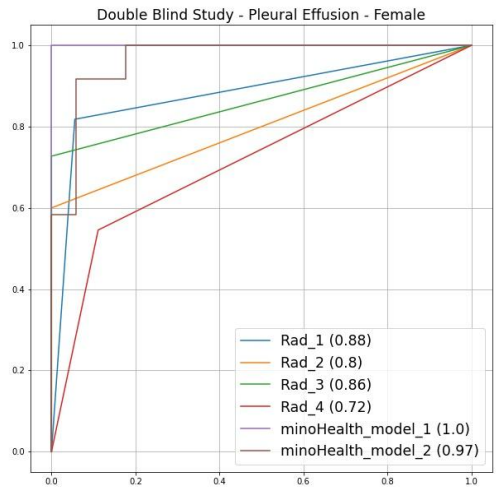

**Figure 4: ROC curves for AI models and radiologists tested on females with cardiomegaly.**

**Figure 6: ROC curves for AI models and radiologists tested on females with pleural effusion.**

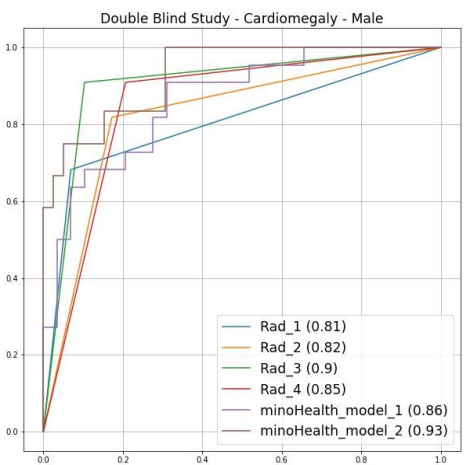

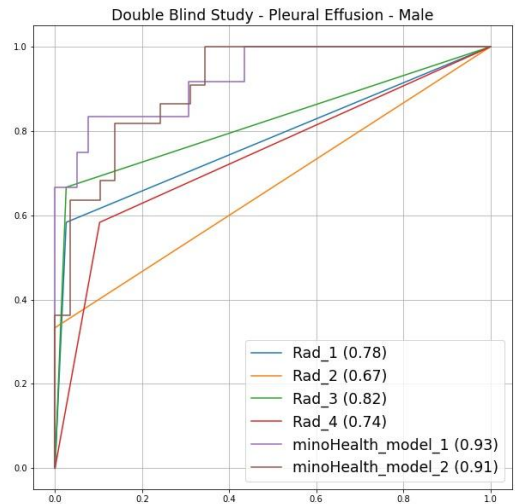

**Figure 5: ROC curves for AI models and radiologists tested on males with cardiomegaly.**

**Figure 7: ROC curves for AI models and radiologists tested on males with pleural effusion.**

For cardiomegaly, the models scored AUC-ROC of 0.94 and 1.0 on female chest x-rays (Figure 4) while they scored 0.86 and 0.93 on male chest x-rays (Figure 5). For pleural effusion, the models scored AUC-ROC of 0.97 and 1.0 on female chest x-rays (Figure 6) while they scored 0.91 and 0.93 on male chest x-ray (Figure 7). Despite the fact the models slightly underperformed on male chest x-rays, they outperformed all radiologists on both gender demographics (even for male chest x-rays) for both conditions. The slight underperformance on the male group could be attributed to some anatomical difference between male and female chest x-rays (e.g. presence of breast tissues).

Country-wise, both models perform very well (> 0.9 AUC-ROC) on chest x-rays from Ghana and USA (Figures 8-11). On the Vietnam data, minoHealth_model_1 scored 0.76 (Figure 12) for cardiomegaly and minoHealth_model_2 scored 0.83 (Figure 13) for pleural effusion. Though these performances are slightly lower than the AI models' average performance, they either outperform or perform comparably to all radiologists.

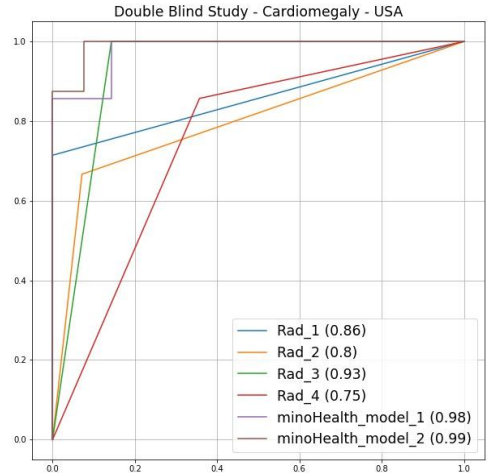

Figure 10: ROC curves for AI models and radiologists tested on American cardiomegaly samples.

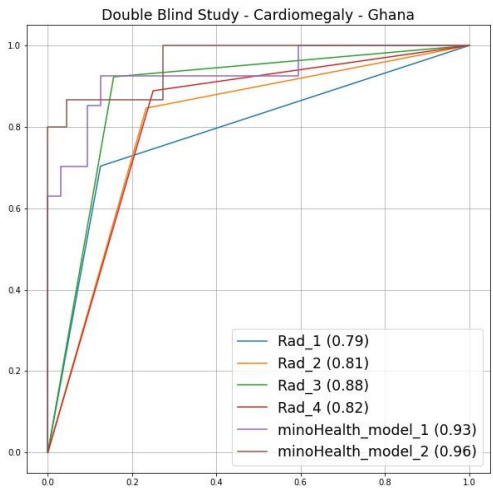

Figure 8: ROC curves for AI models and radiologists tested on Ghanaian cardiomegaly samples.

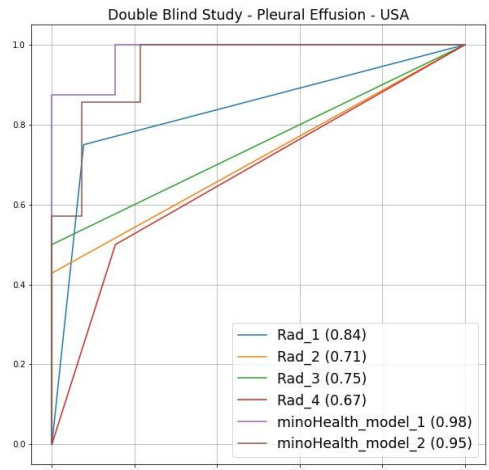

Figure 11: ROC curves for AI models and radiologists tested on American pleural effusion samples.

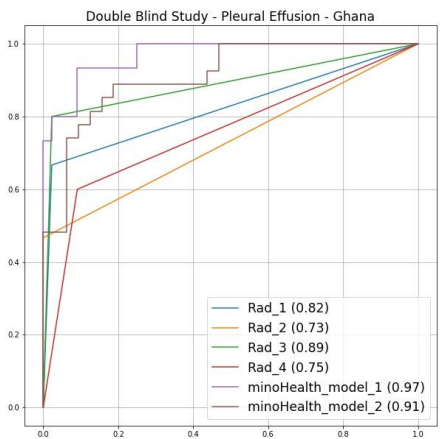

Figure 9: ROC curves for AI models and radiologists tested on Ghanaian pleural effusion samples.

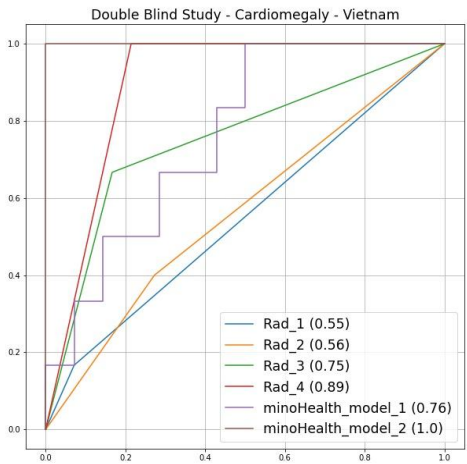

Figure 12: ROC curves for AI models and radiologists tested on Vietnamese cardiomegaly samples.

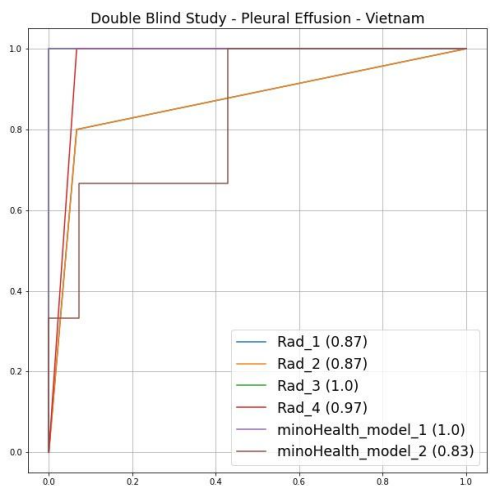

Figure 13: ROC curves for AI models and radiologists tested on Vietnamese pleural effusion samples.

Figures 14 and 15 show the comparison of specificity, sensitivity, negative predictive value (NPV), and positive predictive value (PPV) between AI models and radiologists. Across all four metrics for pleural effusion, either both or one of the AI models outperform the radiologists (Figure 14). A similar observation was made for cardiomegaly, except for sensitivity. For sensitivity on cardiomegaly, minoHealth_model_1 severely underperformed minoHealth_model_2 and other radiologists (Figure 15).

Overall, comparing the performance of the AI models and radiologists across specificity, sensitivity, NPV, and PPV for both cardiomegaly and pleural effusion reinforces the claim that AI models could be used to complement the efforts of radiologists to improve the diagnosis of chest conditions from chest x-ray images.

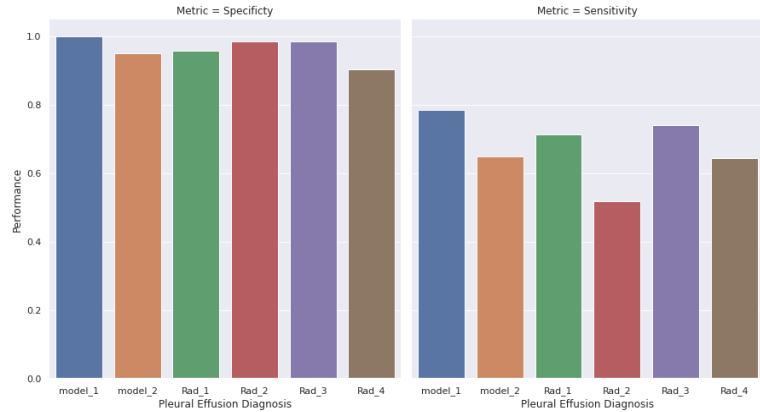

Figure 14a: Specificity and Sensitivity of AI models and radiologists tested on pleural effusion samples.

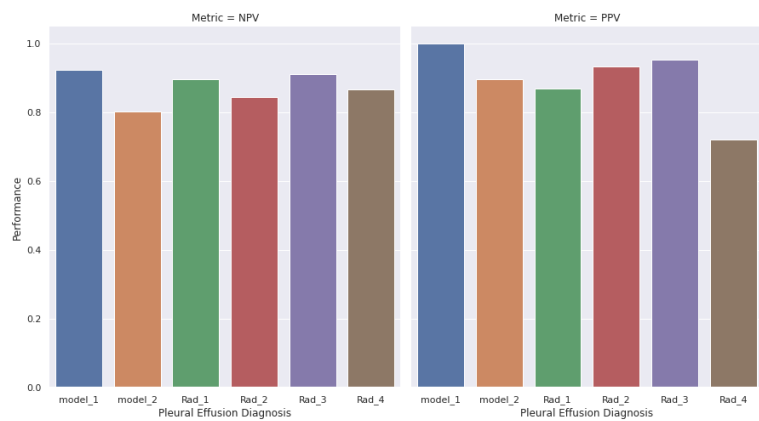

Figure 14b: NPV and PPV of AI models and radiologists tested on pleural effusion samples.

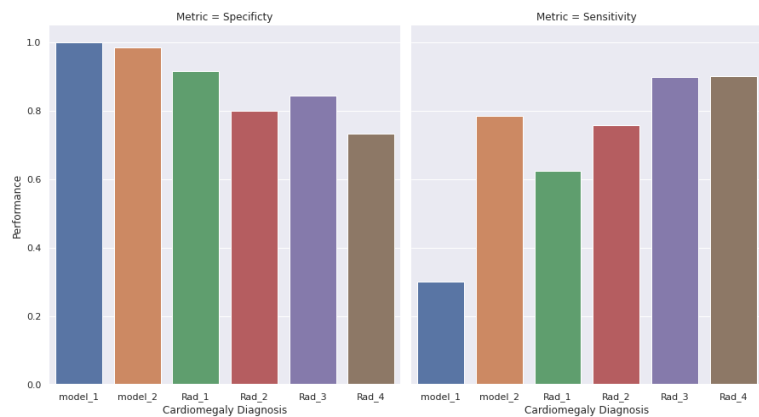

**Figure 15a: Specificity and Sensitivity of AI models and radiologists tested on cardiomegaly samples.**

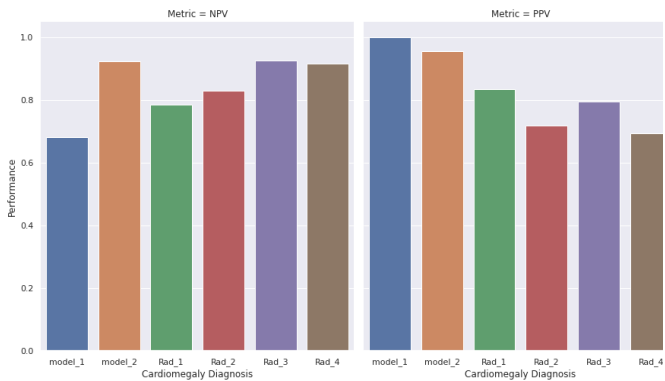

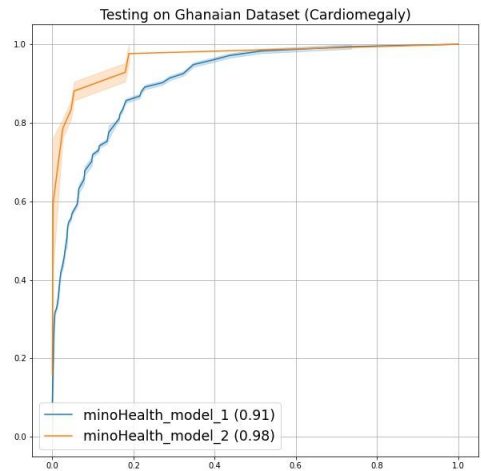

**Figure 15b: NPV and PPV of AI models and radiologists tested on cardiomegaly samples.**

**Figure 16: ROC curves for AI models tested on a larger Ghanaian dataset for cardiomegaly.**

The models were further tested on a larger Ghanaian dataset collected from Euracare Advanced Diagnostics and Heart Centre. The data contained 561 samples. It contained 87 cardiomegaly positive samples, 21 pleural effusion positive samples, and 465 healthy samples. The models were tested on both the full version of the dataset (that contained 87 cardiomegaly samples, 21 pleural effusion samples, and 465 healthy samples) and a more balanced version of the dataset (that contained 87 cardiomegaly samples, 21 pleural effusion samples, and 150 healthy samples).

For cardiomegaly, the models scored AUC-ROC of 0.91 and 0.98 (Figure 16) on the full dataset and scored 0.90 and 0.97 respectively on the balanced version dataset (Figure 18). For pleural effusion, the models scored AUC-ROC of 0.97 and 0.92 (Figure 17) on the full dataset and scored 0.96 and 0.90 (Figure 19) respectively on the balanced version of the dataset.

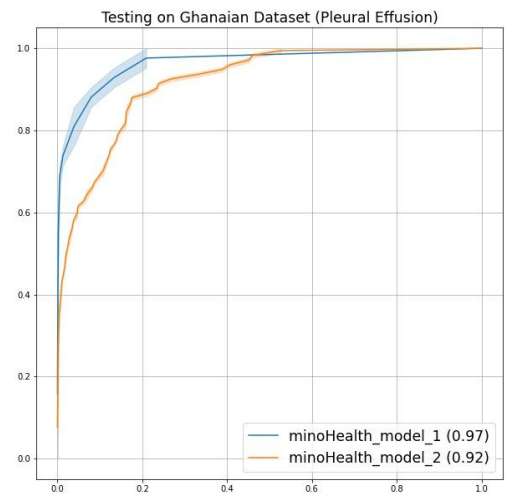

**Figure 17: ROC curves for AI models tested on a larger Ghanaian dataset for pleural effusion.**

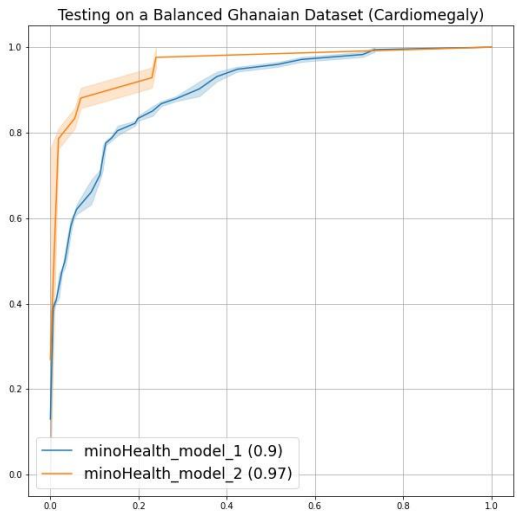

Figure 18: ROC curves for AI models tested on a balanced version of the Ghanaian dataset for cardiomegaly.

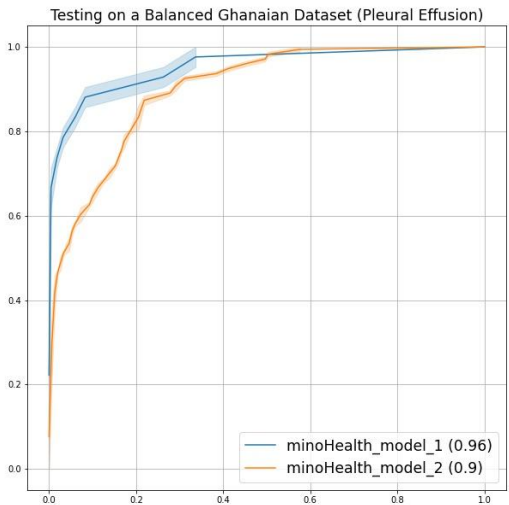

Figure 19: ROC curves for AI models tested on a balanced version of the Ghanaian dataset for pleural effusion.

and cardiomegaly. For both conditions, the best performing AI model outperforms the best performing radiologist by about 10%. These models will be of great utility in regions, like Sub-Saharan Africa, where there are scarce radiologists. To further ascertain the utility and performance of the AI models in regions like Ghana, the models were evaluated on a larger Ghanaian dataset. These AI models can potentially be used to augment the effort of radiologists to improve the diagnosis and treatment of chest conditions.

## 4. CONCLUSION

In conclusion, this study compared the performance of two minoHealth.ai's models to radiologists on a dataset that featured samples from Ghana, USA, and Vietnam. The study focused on two chest conditions - pleural effusion